\documentclass[aps,prb,twocolumn,showpacs,a4paper,superscriptaddress]{revtex4-2}
\usepackage{graphicx,bm,amsmath,dcolumn,amssymb}
\usepackage{newtxtext,newtxmath}
\usepackage{caption}
\usepackage{subcaption}
\usepackage{tikz,pgfplots}
\usepackage[colorlinks]{hyperref}
\hypersetup{
  linkcolor = {blue},
  citecolor = {blue},
  urlcolor = {cyan}
}
\graphicspath{{figures}}
\usepackage{graphicx}
\usepackage{geometry,hyperref,color}
\usepackage[normalem]{ulem}
\usepackage{xpatch}
\newcommand\soutks{\bgroup\markoverwith{\textcolor{blue}{\rule[0.2ex]{2pt}{2.4pt}}}\ULon}
\newcommand\scalemath[2]{\scalebox{#1}{\mbox{\ensuremath{\displaystyle #2}}}}

\begin{document}
	\title{Dynamics of the Baxter-Wu model}
	\author{Chen Tang}
	\affiliation{School of Microelectronics $\&$ Data Science, Anhui University of Technology, Maanshan 243002, China }
	\author{Konstantinos Sfairopoulos}
	\email{kosfairo@gmail.com}
	\affiliation{School of Physics and Astronomy, University of Nottingham, Nottingham, NG7 2RD, United Kingdom}
	\affiliation{Centre for the Mathematics and Theoretical Physics of Quantum Non-Equilibrium Systems, University of Nottingham, Nottingham, NG7 2RD, United Kingdom}
	\author{Wanzhou Zhang}
	\affiliation{College of Physics, Taiyuan University of Technology, Taiyuan 030024, China}
	\author{Chengxiang Ding}
	\email{dingcx@ahut.edu.cn}
	\affiliation{School of Microelectronics $\&$ Data Science, Anhui University of Technology, Maanshan 243002, China }
	\date{\today}
	\begin{abstract}
		Using Monte Carlo simulations, we investigate the dynamical properties of the Baxter-Wu (BW) model under linear quenches.
		For the linear cooling process, the scaling behavior of the excess defect density in the critical region aligns well with the predictions of the Kibble-Zurek (KZ) mechanism. However, the scaling behavior of the excess defect density after exiting the impulse regime does not follow from a simple interplay between the KZ mechanism and the coarsening dynamics; the system undergoes a decay close to a power-law form with an exponent that is significantly different from the coarsening exponent observed in instantaneous quenching.
		For the linear heating process, we show that if the system starts from its ground state, 
        the relevant exponents describing the KZ mechanism are identical to those in the cooling scenario. We find that the system does not directly enter the adiabatic regime after leaving the impulse regime but instead passes through a crossover regime with an exponential decay of the excess defect density. If the initial state is ordered but not the ground state of the system, the defect density exhibits a good scaling behavior, but the relevant exponents do not conform to the predictions of the KZ mechanism.
		
	\end{abstract}
	\maketitle 
	
\section{Introduction}
Phase transitions and critical phenomena remain fascinating research topics in statistical physics, encompassing studies of both equilibrium states and non-equilibrium dynamical processes such as the KZ mechanism\cite{KZ1,KZ2,KZ3,KZ4,KZ5,KZ6,KZ7,KZ8,KZ9,kze1,kze2,kze3,kze4,kze5,kze6,kze7,2014_Liu}, coarsening processes\cite{cos1,cos2,cos3,cos4,cos5,cos6}, aging phenomena\cite{ag1,ag2,ag3,ag4,ag5,ag6}, and those systems with short-time dynamics\cite{short1, short2,short3,short4,short5,short6,short-bw1,short-bw2} for both classical and quantum systems, as well as ordered and disordered systems \cite{sachdev}. The KZ mechanism has also been experimentally verified in a variety of systems; see, for example, Refs.~\cite{2016_Donadello,2019_Ko,2021_Goo,2022_Goo}.

When a given parameter of a system possessing a continuous thermal phase transition, such as the temperature, is linearly adjusted to cross the given transition point, the time dependence of the temperature can be expressed as $T(t) = T_c(1 - t/\tau_Q)$, with the time starting at $t=-\tau_Q$, reaching the critical point $T_c$ at $t=0$, and dropping to zero at $t = \tau_Q$, similar to the transcritical scenario of Ref.~\cite{2012_Chandran}. According to the KZ theory, the entire driving process can be roughly divided into an impulse regime surrounded by two adiabatic regimes for short and long times \cite{KZ4}. 
During the impulse regime and due to the proximity to the continuous phase transition, the system can no longer adjust adiabatically with time. The critical slowing down dominates, resulting in a large correlation length and a long relaxation time for the system. The theory of critical phenomena predicts that close to the critical point the correlation length scales with temperature as $\xi \sim |T-T_c|^{-\nu}$ and the relaxation time takes the form $ \tau_{_{\rm relax}} \sim |T-T_c|^{-\nu z}$, 
where $\nu$ the correlation length critical exponent, and $z$ the critical dynamical exponent\cite{sachdev,book_Goldenfeld}.

In the adiabatic regime, the relaxation time is relatively short, and the relaxation rate of the system is faster than the driving rate, $1/\tau_Q$. The process can be regarded as quasi-static, with the system being close to thermal equilibrium. The boundary between the impulse regime and the adiabatic regime, which is expressed by the time $\hat{t}$,  can be determined by the condition $d\tau_{_{\rm relax}}/dt = 1$, which translates to:
\begin{eqnarray}
    |\hat{t}|  \sim \tau_Q^{\frac{\nu z}{1+\nu z}}.
\end{eqnarray}
The state at $\hat{t}$ can still be considered as an equilibrium state, at which point the correlation length becomes
\begin{eqnarray}
    \xi[T(\hat{t})]\sim \tau_Q^{\frac{\nu}{1+\nu z}},
\end{eqnarray}
and, subsequently,  the excess defect density (EDD) of the system conforms to the following relationship\cite{KZ3,KZ4}
\begin{eqnarray}
    \delta n(\hat{t})=n(\hat{t})-n^{\rm eq}[T(\hat{t})]\sim	\xi[T(\hat{t})]^{-(D-d)}  = \tau_Q^{-\frac{(D-d)\nu}{1+\nu z}}. \label{kinkscal}
\end{eqnarray}
Here $n^{\rm eq}(\hat{t})$ is the density of topological defects of the equilibrium state at the temperature $T(\hat{t})$;
$D$ is the spatial dimension of the system and $d$ the dimension of the topological defect. The KZ mechanism also applies to quantum phase transitions, with the driving parameter being replaced by a quantum parameter
\cite{KZ5,KZ6,KZ7,KZ8,kze1,kze2,kze3,kze4,kze5}.

If a parameter of the system is instantaneously quenched from high temperature to low temperature, the system will undergo coarsening, during which the EDD of the system follows a growth process described by\cite{cos1}
\begin{eqnarray}
    \delta n(t) \sim t^{-1/z_d},
\end{eqnarray}
where we denote by $z_d$ the dynamical exponent for the quenching to the low-temperature phase. Note that, generally, $z\ne z_d$. Furthermore, the value of $z_d$ for quenches at low temperatures may also differ from quenches to zero temperature \cite{cos2,cos4}.

For a linear quench process in a classical system after the system exits the impulse regime, the subsequent dynamics will be influenced by the coarsening. This may lead to the emergence of novel scaling behaviors for the EDD at the endpoint of the quench process.
For example, Refs. \onlinecite{Ising1}-\onlinecite{Ising2} examined this question for the Ising model. This paper will investigate a similar scenario in the context of the BW model. 

The BW model differs significantly from the Ising model in that it consists of three-spin interactions in both the downward- and upward-pointing triangles of the triangular lattice. Although the BW model was found to have a continuous thermal phase transition at exactly the same temperature as the 2D Ising model \cite{bwexact,baxterbook}, the properties of its excitations are completely different. The KZ mechanism is therefore also adapted to apply to the ``defects'' of the BW model.

Our work encompasses both linear cooling and heating processes, both of which align well with the KZ mechanism, assuming that the initial state for the linear quenches represents an equilibrium one. During the linear cooling process, after exiting the impulse regime, the excess defect density primarily undergoes a nearly power-law decay, with a decay exponent that differs significantly from the coarsening exponent observed for instantaneous quenches. For the linear heating process, the system does not directly enter the adiabatic regime after leaving the impulse regime; instead, the EDD first undergoes an exponential decay. Furthermore, we observed an intriguing phenomenon: if the initial state for the heating is ordered but not the ground state of the system, the defect density also exhibits good scaling behavior, but the relevant exponents do not conform to the predictions of the KZ mechanism.

In Sec.~\ref{introduce the model} we introduce the BW model and the relevant quantities needed for the study of the KZ mechanism in the subsequent sections and in Sec.~\ref{res} we present our results. Specifically, in Sec.~\ref{resA} we study the linear cooling case, in Sec.~\ref{Lheat} the linear heating, and in Sec.~\ref{anomalous heating} the anomalous heating case. Lastly, in Sec.~\ref{conclusions} we give our conclusions.

\section{Model and Method}
\begin{figure}
    \centering
    \begin{subfigure}[b]{0.495\columnwidth}
        \includegraphics[width=4.4cm]{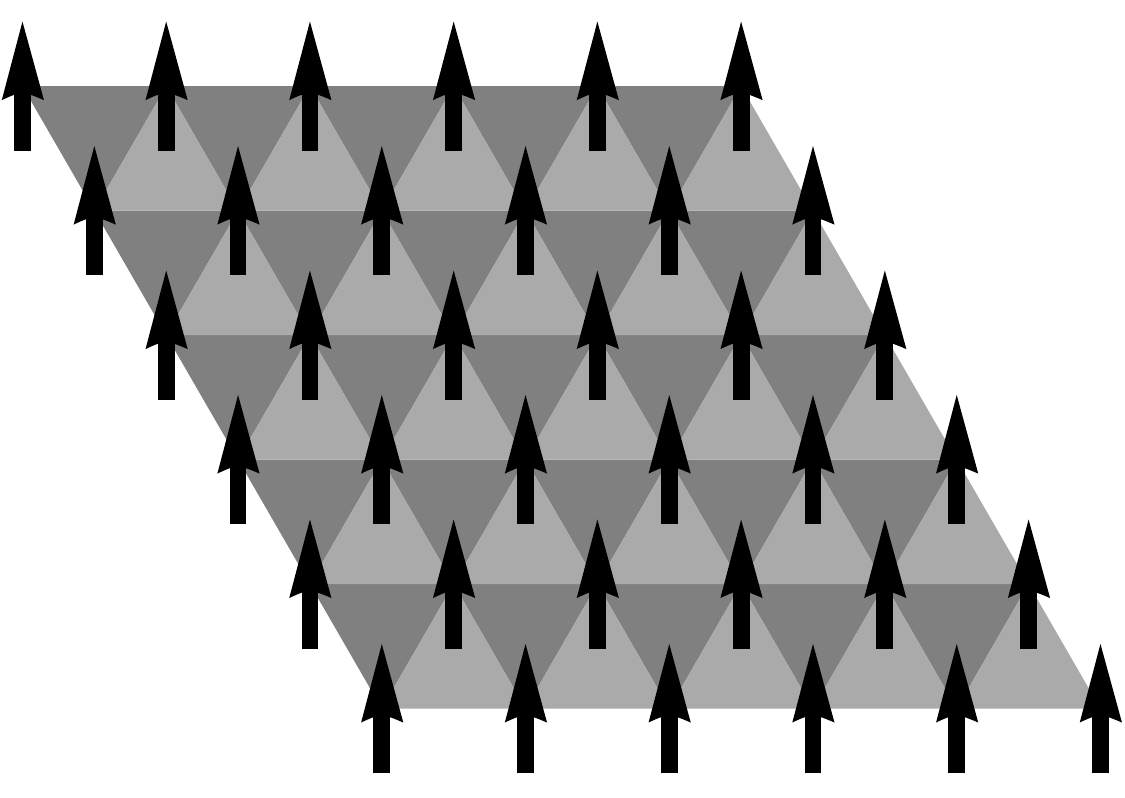}
    \end{subfigure}
    \begin{subfigure}[b]{0.495\columnwidth}
        \includegraphics[width=4.4cm]{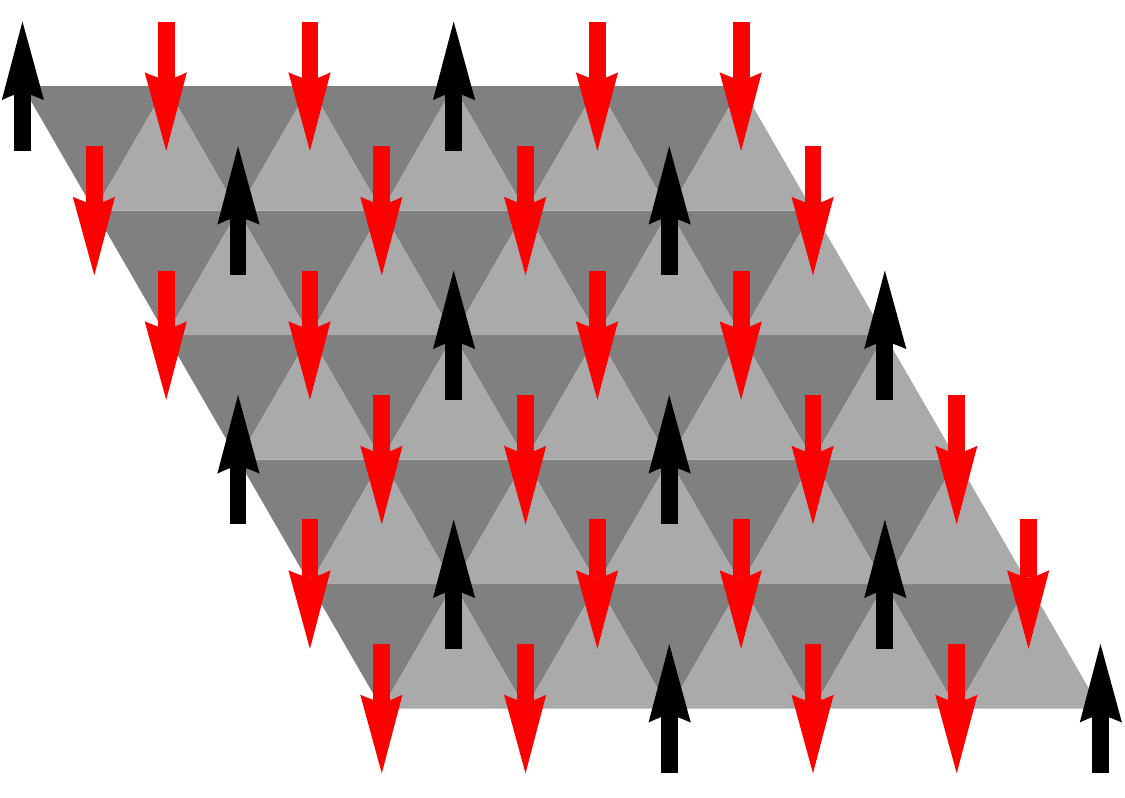}
    \end{subfigure}
    \caption{(Color online) Ground states of the BW model.}
    \label{gs}
\end{figure}

{\label{introduce the model}}
\label{model}
The Baxter-Wu (BW) model was first introduced as an exactly solvable model in statistical physics in Ref.~\cite{bwexact}. It is defined on a two-dimensional triangular lattice, with its Hamiltonian taking the form
\begin{eqnarray}{\label{BWmodel}}
    E=-J\sum\limits_{\triangledown/\triangle}S_iS_jS_k,
\end{eqnarray}
where $S_i=\pm 1$ represents the (classical) spin degree of freedom at site $i$, and $\{i, j, k\} \in {\triangledown/\triangle}$ the three spins on each triangular plaquette of the triangular lattice. The lattice has size $N = L \times L$ and the sum runs over all $2N$ triangular interaction terms of each type. 
As shown in Fig. \ref{gs}, the model has four zero-temperature ground states for open boundary conditions (OBC); one has all spins up ($S_i =+1$) and magnetization $m$=1 and the other three have $S_i=+1$ on only one of the three triangular sublattices, while for the other two $S_i = -1$ and thus $m=-1/3$. The zero-temperature ground states of the model can also be found by the simultaneous evolution of two counterpropagating one-dimensional cellular automata \cite{2023_Sfairopoulos}.
For periodic boundary conditions (PBC), the $m=-1/3$ states are ground states of the model only when both linear dimensions, $L$, are an integer multiple of 3. 

The phase transition of the model belongs to the 4-state Potts universality class, with critical exponents $ \nu= 2/3$ and $\beta= 1/12$.
An interesting point to note is that this model does not exhibit the logarithmic corrections\cite{bwgd} that are commonly found in the 4-state Potts universality class\cite{log}. The exact solution provided by Baxter and Wu shows that in the thermodynamic limit the partition function of the model, $Z$, satisfies the relation \cite{bwexact}
\begin{eqnarray}
    W=\lim\limits_{N\rightarrow\infty}Z^{1/N}=\sqrt{6yt}, \label{WW}
\end{eqnarray}
where $t = \sinh(2K)$, with $K = |J|/k_BT$, and $1\le y\le\infty$ a root of the equation 
\begin{eqnarray}
    \frac{(y-1)^3(1+3y)}{y^3}=\frac{2(1-t)^4}{t(1+t^2)}.
\end{eqnarray}
The singular point of $\log W$ gives the critical point of the phase transition of the model, $k_BT_c/|J| = 2/\log(1 + \sqrt{2})$.

The critical dynamical properties of the BW model have also been studied using the method of short-time dynamics \cite{short-bw1}, which revealed a critical dynamical exponent $z\approx2.294$ for the model with Glauber dynamics. Additionally, if the system is instantaneously quenched to its critical point starting from an initial state that is ordered but not one of the model's ground states, the model exhibits anomalous (short-time) dynamical behavior \cite{short-bw2}.
	
In this work, we will focus on the dynamical properties of the model under linear quenches. The physical quantities we aim to calculate include the EDD, $\delta E(t)$, and the  squared magnetization, $m^2(t)$, which are defined as follows
\begin{eqnarray}
    \delta E(t)&=&E(t)-E^{\rm eq}[T(t)], \label{deltaE}\\
    m^2(t) &=&\Big\langle \Big(\frac{1}{N}\sum\limits_{i}S_i(t)\Big)^2\Big\rangle,\label{msub}
\end{eqnarray}
with 
\begin{eqnarray}
    E(t)=\Bigg\langle \frac{1}{2N}\sum\limits_{\{i,j,k\}}^{2N} \frac{1-S_i(t)S_j(t)S_k(t)}{2}\Bigg\rangle. \label{Et}
\end{eqnarray}
Here, $\langle \cdots \rangle$ denotes the average over different initial states
and $E(t)$ is the defect density (DD) of the nonequilibrium state, while
$E^{\rm eq}[T(t)]$ is the DD of the equilibrium state at the corresponding temperature.
For the BW model, we define a defect as an energy excitation of the smallest unit, specifically $S_iS_jS_k =-1$.
 $E^{\rm eq}[T(t)]$  is related to the internal energy density $U$ by
\begin{eqnarray}
    E^{\rm eq}[T(t)]=0.5+\frac{U}{4}
\end{eqnarray}
with $U$ given by the formula
\begin{eqnarray}
    U=T^2\frac{\partial}{\partial T}(\log W).
\end{eqnarray}
We set $k_B=1$ and $J=1$ throughout this work.
	
For continuous phase transitions, the specific heat scales as $C_V \sim | T - T_c|^{-\alpha}$. Taking the reasonable assumption that the hyperscaling relation $\alpha = 2 - \nu D$ holds below the upper critical dimension of the model \cite{book_Cardy,book_Goldenfeld}, the energy at the critical point scales as $E(T) - E(T_c) \sim |T - T_c|^{\nu\Delta_E}$, where the scaling dimension of the primary field of the underlying CFT is given by $\Delta_E= D - 1/\nu$. When studying the KZ mechanism, we need to take this scaling dimension into account, 
meaning that Eq.~\ref{kinkscal} evaluates to
\begin{eqnarray}
    \delta E(\hat{t})\sim	\xi[T(\hat{t})]^{-\Delta_E}  = \tau_Q^{-\frac{\Delta_E\nu}{1+\nu z}}. \label{Escal}
\end{eqnarray}
From Eqs.~\ref{kinkscal} and \ref{Escal}, we then relate $d$, the dimension of the topological defect, to $d = 1/ \nu$
with $d=1$ for the domain walls of the two-dimensional classical ferromagnetic Ising model\cite{Ising1} and $d=3/2$ for the BW model, since it belongs to the 4-state Potts universality class.

In the following, we will numerically study the KZ mechanism for the BW model and
also investigate the scaling behavior of the EDD, $\delta E(t)$, at the cooling endpoint, which requires consideration of the real-time dynamics after $\hat{t}$. In Ref.~\onlinecite{Ising1}, for the case of the Ising model, the authors adopted the following scaling form
\begin{eqnarray}
    \delta n(\tau_Q) \sim \tau_Q^{-\frac{\nu}{1+\nu z}}\mathcal{F}(t/\hat{t})
\end{eqnarray}
for $t/\hat{t} \gg1$. The function $\mathcal{F}$ should conform to the standard coarsening, i.e., $\mathcal{F}(t/\hat{t})\sim (t/\hat{t})^{-1/z_d}$. Therefore, for $t = \tau_Q$ (assuming $\tau_Q$ is sufficiently large), the EDD should conform to the following scaling form
\begin{eqnarray}
\delta n(\tau_Q) \sim \tau_Q^{-\frac{\nu}{1+\nu z}}(\tau_Q/\hat{t})^{-1/z_{_d}}=\tau_Q^{-\frac{1}{z_{_d}}\frac{1+z_{_d}\nu}{1+z\nu}}. \label{kzth}
\end{eqnarray}

In this paper, we will closely examine the real-time dynamical behavior of the BW model following $\hat{t}$, to understand the scaling behavior of the excess defect density at $t = \tau_Q$. The results obtained are significantly different from those for the Ising model, as detailed in Sec. \ref{resA}.

Our simulation method is the Glauber dynamics, where the spin at lattice site $i$ flips from $S_i$ to $-S_i$ with the following probability \cite{glauberDyn}
\begin{equation}
    \scalemath{0.96}{
        P(S_i\rightarrow-S_i)=\frac{1}{2}\Big\{1-S_i\tanh\Big[\frac{J}{k_BT}\Big(\sum\limits_{k=1}^6S_i^{(k)}S_i^{(k+1)}\Big)\Big]\Big\}
    }. \label{glauber}
\end{equation}
Here, $S_i^{(k)}$ corresponds to the $k$-th nearest-neighbor spin of $S_i$ with $S_i^{(7)}\equiv S_i^{(1)}$. For our simulation, we use $L=600$ and PBC and the values at each time step are averaged over 200 runs. For cooling scenarios, each run starts with a different initial state at thermal equilibrium; for heating scenarios, the initial state is the same, but each run employs a different random number seed for the dynamics.
For the linear heating case, we consider not only the heating starting from the ground states but also from the states with $m=-1$ or $m=1/3$, which are obtained by a global spin-flip of the ground state configurations.

\section{results}
\label{res}
\subsection{Linear cooling}
\label{resA}
\begin{figure}[thpb]
    \includegraphics[width=0.95\columnwidth]{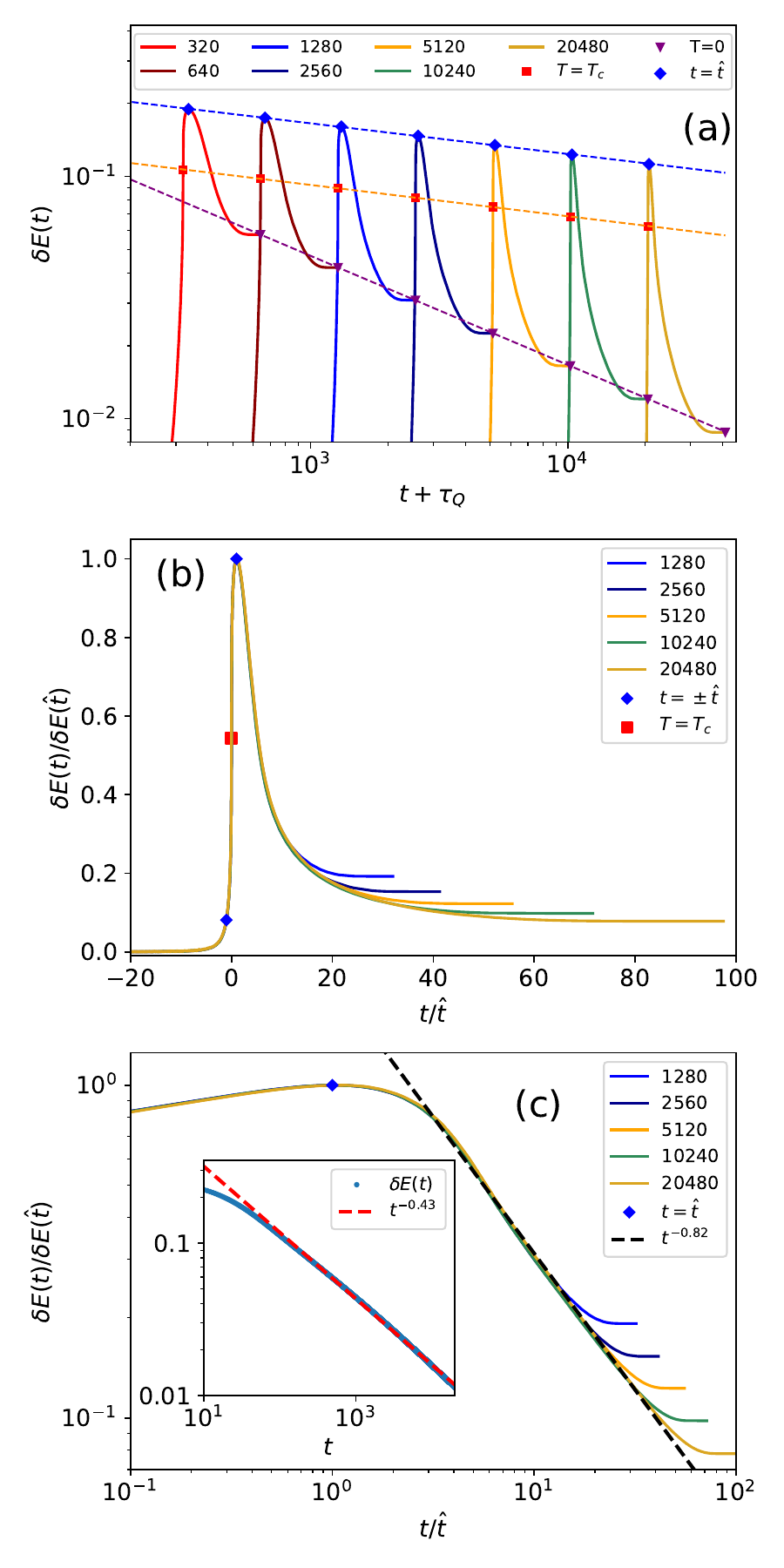}
    \caption{(Color online) (a) Dynamics of the EDD of the BW model under linear cooling according to Eq.~\ref{ramp}, 
        with an initial time  $t=-\tau_Q$. The numerical values of the legend denote the values of $\tau_Q$ for different quenches.
        Note that, in order to better reveal the scaling behavior at times $t=0$ ($T=T_c$), $t=\hat{t}$, and $t=\tau_Q$ ($T=0$), we have shifted the time coordinate by $\tau_Q$. (b) Data collapse of $\delta E(t)/\delta E(\hat{t})$ versus $t/\hat{t}$ for different $\tau_Q$.  (c) Similar to panel (b) but in a logarithmic scale. 
        Inset: relaxation of the EDD following an instantaneous quench from a high-temperature equilibrium state to $T = T_c/2$.}
    \label{eed}
\end{figure} 
We linearly cool the BW model according to the ramp 
\begin{eqnarray}
    T(t) = T_c(1 - t/\tau_Q), \label{ramp}
\end{eqnarray}
with an initial temperature $T = 2T_c$ at $t = -\tau_Q$,
critical temperature $T = T_c$ at $t = 0$,  and zero temperature $T = 0$ at the ending point of the cooling process at $t = \tau_Q$. Fig.~\ref{eed}(a) presents the plot of the EDD as a function of time, where the data points at \(t = 0\), \(t = \hat{t}\), and \(t = \tau_Q\) have been specially marked. The data at these three times exhibits excellent scaling, 
\begin{eqnarray}
    \delta E(t=0) &\sim& \tau_Q^{-0.132\pm 0.002} \label{cooln1}\\
    \delta E(t=\hat{t}) &\sim& \tau_Q^{-0.131\pm 0.001}\label{cooln2}\\
    \delta E(t=\tau_Q) &\sim& \tau_Q^{-0.452\pm 0.002}. \label{EtauQ}
\end{eqnarray}
It can be seen that the exponents of the data in \(t = 0\) and \(t = \hat{t}\) agree well with the theoretical value $\Delta_E\nu/(1+\nu z)=0.132$ given by Eq.~\ref{Escal}, where $\nu=2/3$, $z=2.294$, and $\Delta_E=0.5$. Fig.~\ref{eed}(b) shows the data collapse for $\delta E(t)/\delta E(\hat{t})$ versus $t/\hat{t}$. In the impulse regime $[-\hat{t}, \hat{t}]$, the scaling of the data further demonstrates that the EDD for the BW model conforms to the predictions of the KZ mechanism, cf. Eq.~\ref{Escal}.

To understand the scaling behavior at $t = \tau_Q$, we carefully analyzed the real-time dynamical behavior following $\hat{t}$. As shown in Fig.~\ref{eed}(c), the process after $\hat{t}$ is primarily dominated by a power-law decay with an effective critical dynamical exponent   
\begin{eqnarray}
    \delta E(t)\sim t^{-1/z_{\rm eff}}\approx t^{-0.82} \quad (t>\hat{t}), \label{afterthat}
\end{eqnarray}
with $z_{\rm eff} = 1.220(5)$. At $t = \tau_Q$,
\begin{equation}
    \scalemath{0.94}{
        \delta E(\tau_Q) \sim \tau_Q^{-\frac{\Delta_E\nu}{1+\nu z}}(\tau_Q/\hat{t})^{-1/z_{\rm eff}}=\tau_Q^{-\frac{1}{z_{\rm eff}}\frac{1+\Delta_Ez_{\rm eff}\nu}{1+z\nu}}=\tau_Q^{-0.456}
    }, \label{endscale}
\end{equation}
which explains why the EDD exhibits a decay with a power of $\tau_Q^{-0.452}$ at $t = \tau_Q$, as shown in Eq.~\ref{EtauQ}. It is important to note that the decaying process after $\hat{t}$ does not correspond to a standard coarsening process. For comparison, the inset of Fig. \ref{eed}(c) illustrates the relaxation process from a high-temperature equilibrium state instantaneously quenched to $T = T_c/2$, 
which exhibits a decaying  of 
\begin{eqnarray}
    \delta E(t)\sim t^{-1/z_d}=t^{-0.43\pm 0.02}. \label{sudd}
\end{eqnarray}
This defines a dynamical exponent $z_d = 2.33(1)$, significantly different from $z_{\rm eff}$.

\subsection{Linear heating}
\label{Lheat}
\begin{figure}[thpb]
    \centering 
    \includegraphics[width=0.85\columnwidth]{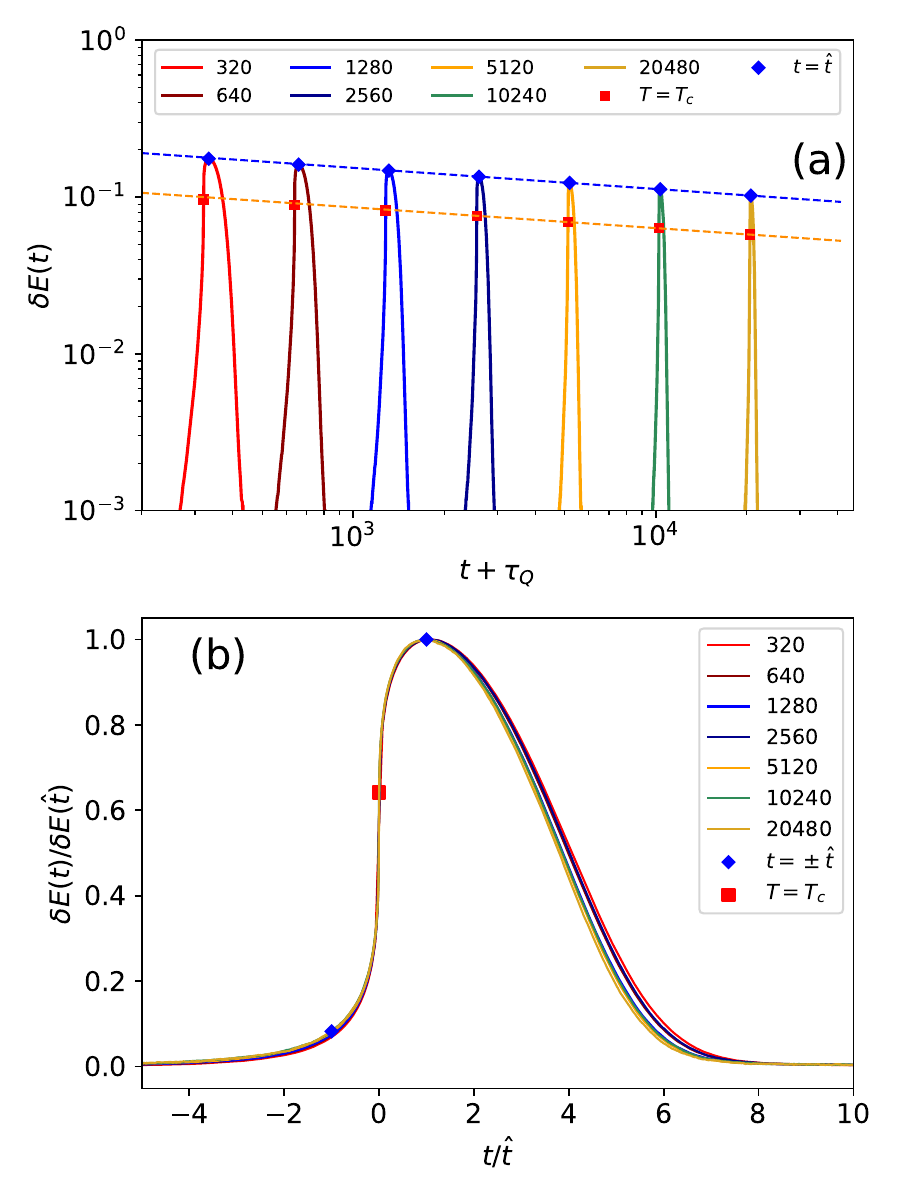}
    \caption{(Color online) (a) Dynamics of the EDD for the linear heating protocol Eq.\ref{heatramp}, similar to Fig.~\ref{eed}a. (b) Data collapse of $\delta E(t)/\delta E(\hat{t})$ versus $t/\hat{t}$ for different $\tau_Q$.}
    \label{eedheat}
\end{figure} 

\begin{figure}[thpb]
    \centering 
    \includegraphics[width=0.85\columnwidth]{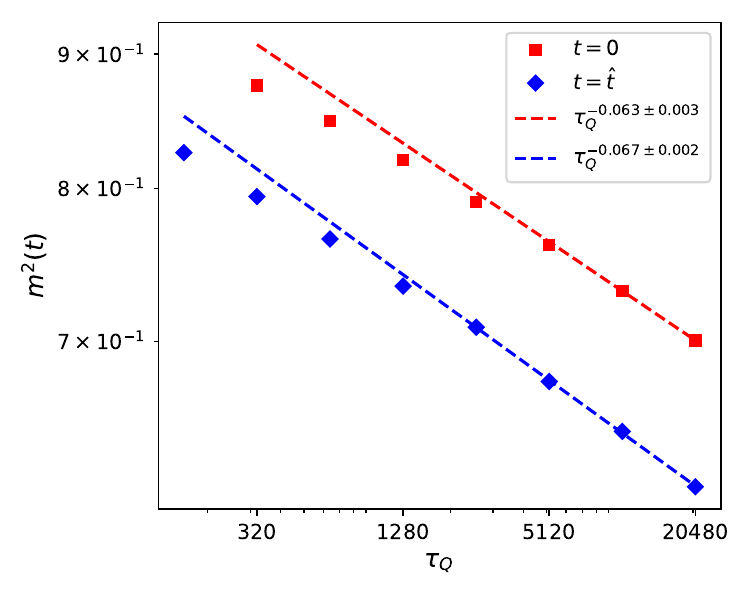}
    \caption{(Color online)  Scaling of the squared magnetization at $t=0 $ ($T=T_c$) and $t=\hat{t}$ for the linear heating of the BW model.}
    \label{magheat}
\end{figure}

We linearly increase the temperature of the BW model according to the ramp
\begin{eqnarray}
    T(t) = T_c (1 + t/\tau_Q), \label{heatramp}
\end{eqnarray}
with an initial temperature $T=0$ at $t = -\tau_Q$, $T = T_c$ at $t = 0$ and $T=2T_c$ for $t = \tau_Q$.  
The initial state is the ground state of the system. As mentioned earlier, the system has two types of ground states with $m=1$ and $m=-1/3$. 
Here, we only present the results for the case where $m(t=0)=1$. The results for the case where $m(t=0)=-1/3$ are qualitatively the same as those for $m(t=0)=1$.
	
Fig.~\ref{eedheat}(a) presents the plot of the EDD of the heating process as a function of time, 
where data points at $t = 0$ and $t = \hat{t}$ have been specially marked. 
Similarly to Eqs.~\ref{cooln1} and \ref{cooln2}, we find that
\begin{eqnarray}
    \delta E(t=0) &\sim& \tau_Q^{-0.130\pm 0.002}, \label{heatn1}\\
    \delta E(t=\hat{t}) &\sim& \tau_Q^{-0.132\pm 0.001}. \label{heatn2}
\end{eqnarray}
It is apparent that the exponents agree well with the theoretical value of $0.132$, given by Eq.~\ref{Escal}. 
They also coincide with the numerical results for the cooling process, as shown in Eqs.~\ref{cooln1}-\ref{cooln2}.
However, the value of $\delta E(t)$ at $t=\tau_Q$ is already very small and does not show any scaling behavior.
Fig.~\ref{eedheat}(b) shows the data collapse for $\delta E(t)/\delta E(\hat{t})$ as a function of $t/\hat{t}$. It can be seen that, within the impulse regime $[-\hat{t},\hat{t}]$, the data for different $\tau_Q$ values collapse onto a single curve, further indicating that the EDD of the BW model during the linear heating process aligns well with the predictions of the KZ mechanism.
Furthermore, it can be observed that after $\hat{t}$, within a certain regime, the curves corresponding to different values of $\tau_Q$ do not
follow the same scaling principle. We will discuss this regime in more detail later.

Furthermore, we study the squared magnetization at $t = 0$ and $t = \hat{t}$ and show that it also conforms to the predictions of the KZ mechanism. Similar to the behavior of the EDD, the scaling behavior of the squared magnetization should follow the equation
\begin{eqnarray}
    m^2(\hat{t})\sim	\tau_Q^{-\frac{2\Delta_S\nu}{1+\nu z}}. \label{mscal}
\end{eqnarray}
where $\Delta_S=\beta/\nu = 1/8$ the primary operator of the field $S$\cite{log}.
For the BW model the squared magnetization obeys the relation
\begin{eqnarray}
    m^2(\hat{t})\sim \tau_Q^{-0.0659}.
\end{eqnarray}

Fig.~\ref{magheat} presents numerical results for the squared magnetization at $t = 0$ and $t = \hat{t}$. We observe that at $t = \hat{t}$ the numerical results are in agreement with the theoretical prediction.  At $t = 0$ (the critical point $T=T_c$), the numerical results are also in line with the theory, but show stronger corrections for small values of $\tau_Q$.

\begin{figure}[thpb]
    \centering 
    \includegraphics[width=0.85\columnwidth]{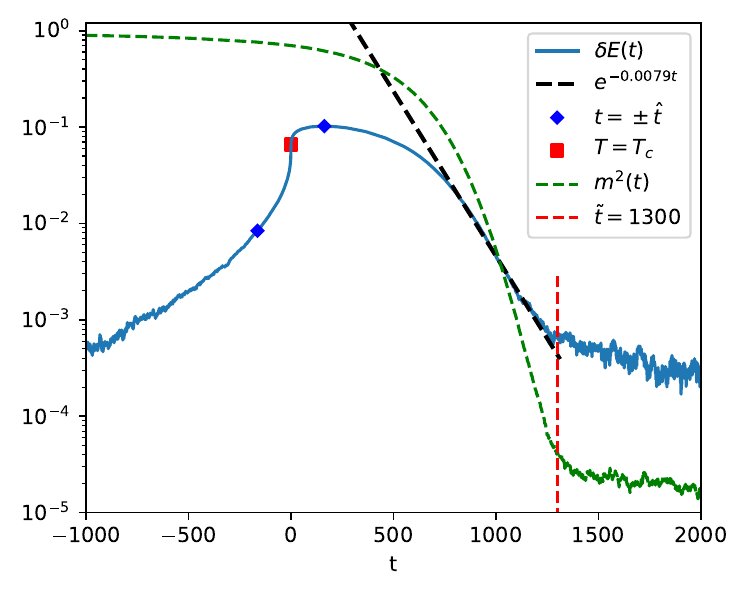}
    \caption{(Color online)  Dynamics of the EDD for the linear heating process after the impulse regime with $\tau_Q=20480$.}
    \label{ertheat}
\end{figure} 
For the process after $\hat{t}$, we discuss the case with $\tau_Q = 20480$ as an example; as shown in Fig.~\ref{ertheat}, 
there is a short regime after $\hat{t}$ during which the EDD decreases exponentially until $\tilde{t}\approx1300$.
After this point, the EDD becomes very small, which means that the DD is approaching the equilibrium value. At the same time, the squared magnetization also becomes very small. After $\tilde{t}$,  the system undergoes a quasi-static process.

The relaxation rate $1/\tau_{\rm eq}$ of the system during the time interval between $\hat{t}$ and $\tilde{t}$ is already faster than the driving rate, but the influence of the driving is still non-negligible. This time regime is still very close to the critical point and thus still affected by its critical behavior. 
In summary, we have found that the transition from the impulse (frozen) region to the adiabatic region is not direct, but is separated by  this regime. As shown in Fig.~\ref{eedheat}(b), during this crossover region, the $\delta E(t)/\delta E(\hat{t})$ as a function of $t/\hat{t}$ does not follow the same scaling principle. The true adiabatic region begins after $\tilde{t}$.

\subsection{Anomalous heating}{\label{anomalous heating}}
\begin{figure}[thpb]
    \centering 
    \includegraphics[width=0.85\columnwidth]{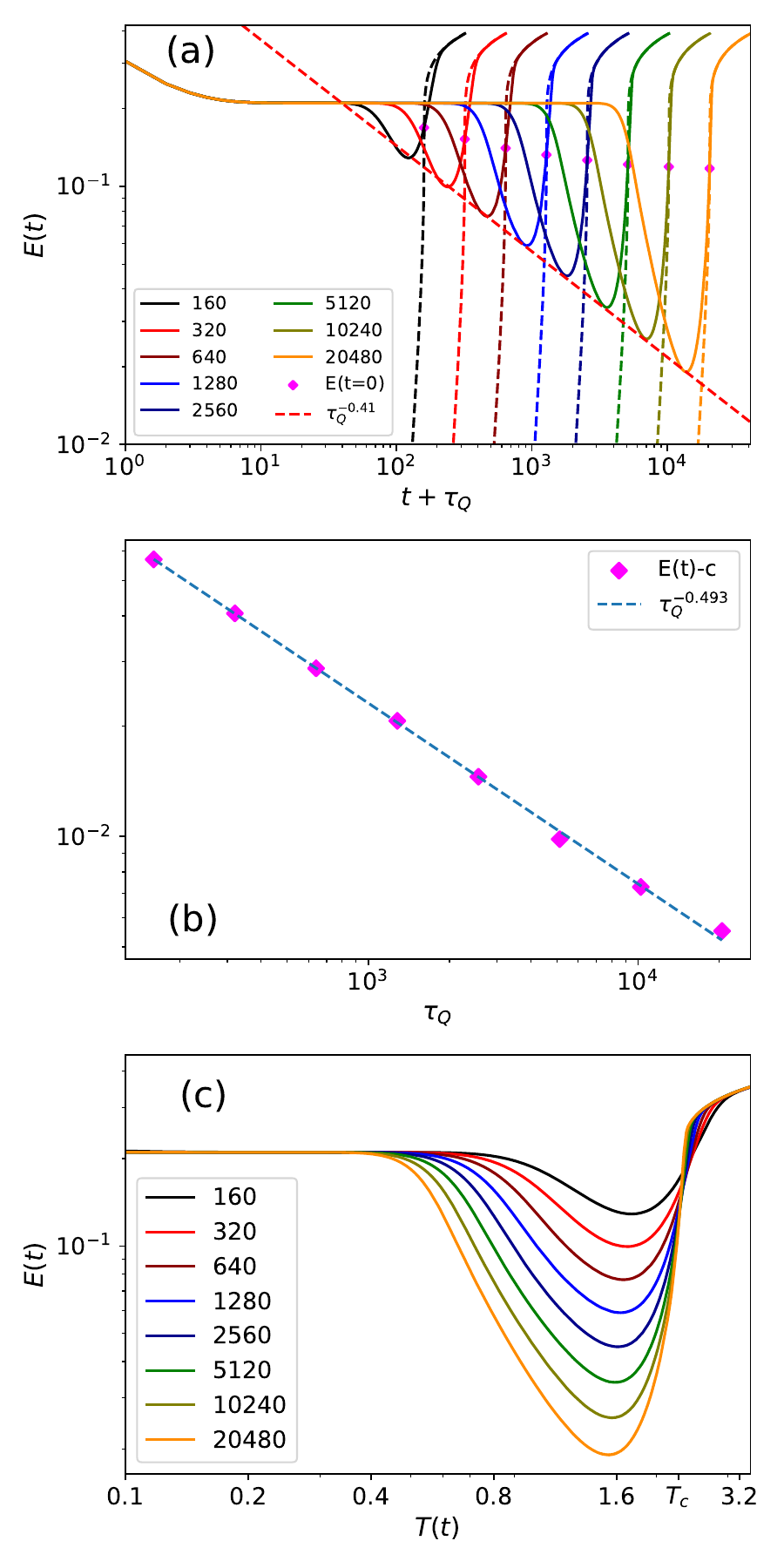}
    \caption{(Color online) (a) Dynamics of the DD of the BW model under linear heating according to Eq.~\ref{heatramp},
        starting from an initial time $t=-\tau_Q$. The legend denotes the values of $\tau_Q$ for different quenches, starting from the initial state with $m=-1$, obtained from the $m=1$ ground state by a global spin flip.
        (b) Scaling of $E(t)$ at $t=0$ ($T=T_c$).
        (c) Replot of the DD, but with the abscissa replaced by $T(t)$.}
    \label{abheat}
\end{figure} 
According to Ref.~\onlinecite{short-bw2}, when instantaneously quenched to the critical point from an initial state that is ordered but not  belonging to the ground state manifold, the system exhibits anomalous short-time dynamical behavior. Inspired by this phenomenon, we consider linear heating protocols from such states. The initial states we used include those states with $m=-1$ and $m=1/3$, which are obtained by a global spin-flip from the ground states. For brevity, only the results for the $m=-1$ state are presented here.

Fig.~\ref{abheat}(a) shows the DD during the linear heating starting from the $m=-1$ state. We observe that, because of the low temperature for the early stages of the protocol, the system undergoes a process of decaying, which apparently is a process dominated by coarsening. After reaching a minimal point, the DD gradually increases again because the temperature continues to increase. The value of the DD at this minimal point scales with $\tau_Q$ as
\begin{eqnarray}
    E_{\rm min}\sim \tau_Q^{-0.41\pm 0.01}.\label{E1}
\end{eqnarray} 
Interestingly, the exponent matches the exponent of Eq.~\ref{sudd}.

Another intriguing phenomenon concerns the DD at the critical point, which we found to decay to a constant as $\tau_Q$ increases, following the relation
\begin{eqnarray}
    E(t) = a\tau_Q^{-b} + c\label{E2},
\end{eqnarray}
where $b = 0.49(1)$ and $c=0.112(1)$. Fig.~\ref{abheat}(b) demonstrates the scaling behavior of $E(t) - c$ with $\tau_Q$. Notably, the exponent $b$ here differs significantly from the exponent given by the KZ mechanism during the normal heating process.
\begin{figure}[thpb]
    \centering 
    \includegraphics[width=0.85\columnwidth]{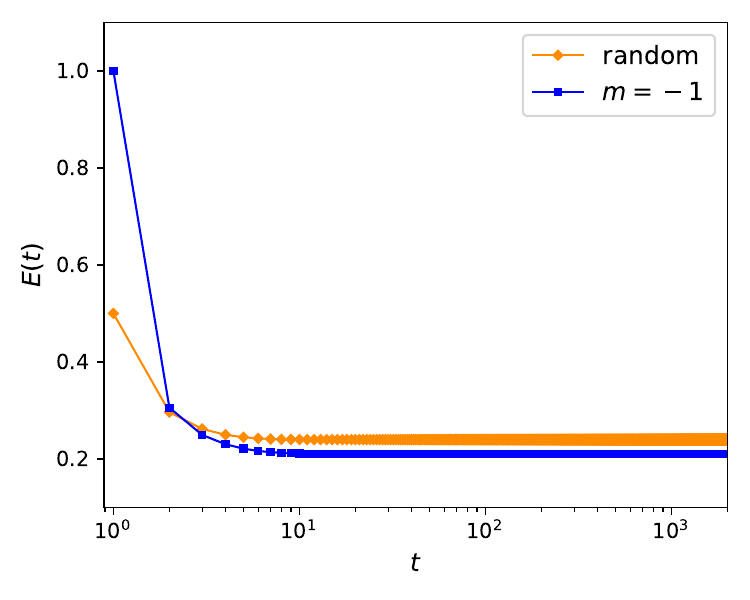}
    \caption{(Color online)  Relaxation of the  energy density following an instantaneous quench from a high-temperature equilibrium state to zero temperature.}
    \label{zero}
\end{figure} 

\begin{figure}
    \centering 
    \includegraphics[width=1\columnwidth]{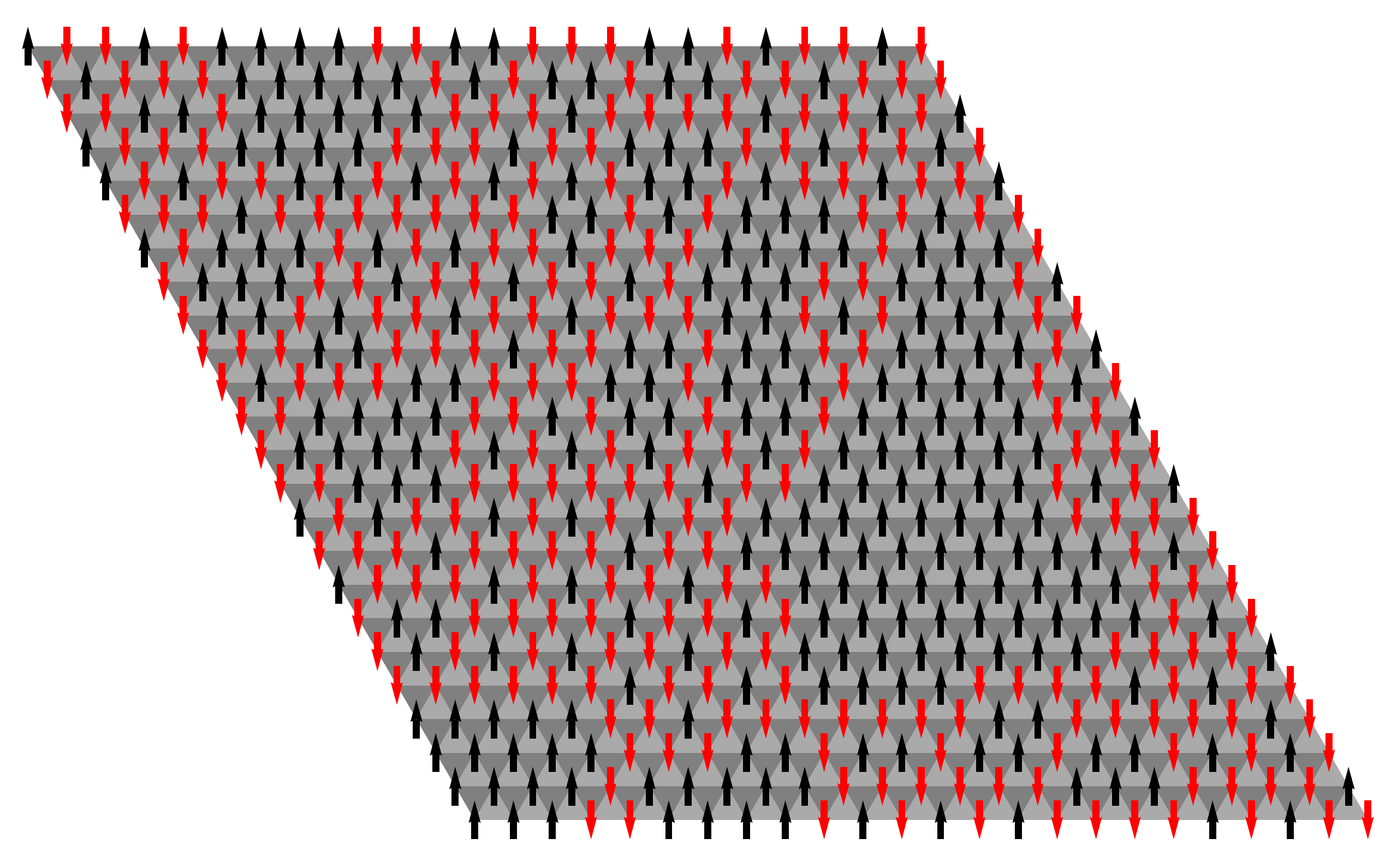}
    \caption{(Color online)  A metastable state of the BW model obtained by an instantaneous quench from high temperature to zero temperature.}
    \label{metas}
\end{figure} 
Regarding the scaling behavior of Eqs.\ref{E1} and \ref{E2}, we currently lack a detailed theoretical explanation. However, through numerical calculations, we have found that this phenomenon is likely related to the special low-temperature behavior the model. 
Fig.~\ref{zero} illustrates the time evolution of the DD after an instantaneous quench from a high-temperature initial state and from an initial state with $m=-1$ to zero temperature.
It can be observed that the DD rapidly decays to a stable value that is not zero (the DD of the ground states must be zero at zero temperature as the model is frustration-free).
In other words, the system relaxes to a metastable state. 
We have also tested quenches to non-zero but small temperatures, and the results were qualitatively consistent with the quenches to zero temperature.

As shown in Fig.~\ref{gs}, the BW model has two types of ground states: the $m=1$ state and the $m=-1/3$ state. When quenched from an initial state with a high number of defects to $T=0$, the system may form a state that is as depicted in Fig.~\ref{metas}. This state is composed of sub-clusters that belong to the $m=1$ and $m=-1/3$ ground states, shown in Fig. \ref{gs}. We have found that at zero or very low temperature, following the Glauber dynamics, the spins in patterns like Fig.~\ref{metas} are very difficult to flip, even those at the boundaries between the different domains. 


Fig.\ref{abheat}a is structured into some distinct regions:
\begin{itemize}
    \item Initially, the DD is very high and the equilibrium DD is zero. 
    \item Following this stage, the DD relaxes rapidly due to the temperature quench.
    \item Gradually, the DD enters a plateau: the DD, cf. Eq.~\ref{deltaE}, doesn't change with the temperature ramp. This is a behavior that is not uncommon in earlier works  \cite{2015_Chesler,2016_Huang,2019_Ko,2023_Kou,2010_Campo,2015_Chesler,2015_Sonner,2019_Gomez,2020_Li,2021_Zeng,2021_Li,2021_Goo,2022_Goo}. However, we argue that the appearance of the plateau in the BW model is different from the above works.
    
    For the BW model, the system after relaxing fast with the temperature gradient, gets quickly trapped in local minima (cf. Fig.~\ref{zero}-\ref{metas}). This part of the quench is similar to an instantaneous quench to zero temperature, since the temperature is still close to zero. The extent of the trapping depends on $\tau_Q$, which is shown in Fig. \ref{abheat}(c), 
    as expected. The increase of $\tau_Q$ for given $t_i$ and $t_f$ gives rise to a slower quench which allows the system to avoid the local energy barriers. The higher the $\tau_Q$ the smaller the plateau and for an infinitely slow temperature gradient the system would theoretically avoid the metastable states. 
    Lastly, the value of the plateau changes depending on the starting high temperature state, which validates our argument for the system being trapped on metastable states. 
    
    \item After the system manages to escape the plateau with increasing temperature, another relaxation is observed. The scaling though is apparently different from the one before the plateau. The relaxation now is much slower than before. For a slow enough linear quench, the system would first equilibrate and the minimum would extrapolate to $\delta E = 0$.
    We expect in this case the KZM to hold. We expect that in the absence of the intermediate plateau, the two relaxation steps would have had the same scaling. Note that our findings for the scaling of the DD, Eq.~\ref{E1}, are consistent with the suppression of defects found in the ordered phase in Ref.~\cite{2015_Chesler}. Note, however, that this region for us is before the phase transition.

    \item After the minimum, as identified by Eq.~\ref{E1}, the system enters the supposedly ``impulse'' regime. However, as it has not managed to equilibrate, the exponents for the critical point differ from the ones of the KZM. We expect the scaling of Eq.~\ref{E2} to gradually approximate the one of Eqs.~\ref{cooln1}, \ref{cooln2}, \ref{heatn1} and \ref{heatn2} for $\tau_Q \rightarrow \infty$. To support our argument, Ref.\cite{2023_Kou} has argued in favor of the existence of a ``pre-saturated'' (PS) regime, before the appearance of the KZ scaling. The extent of this PS regime is system dependent. For the BW model, and from Fig. \ref{abheat}b we start to see the difference in scaling observed for $\tau_Q=20480$ compared to the smaller values, which is evident in the existence of the constant prefactor in Eq.~\ref{E2}. As discussed in Ref.~\cite{2023_Kou} we expect the appearance of a zeroth-order term (cf. Eq.31 of Ref.\cite{2023_Kou}), which corroborates our finding, Eq.~\ref{E2}. By increasing the value of $\tau_Q$ we expect the exponent to approach $0.132$ and the constant prefactor to vanish.
    \item Lastly we have an increase of the defect energy density, as expected, close to and after the critical point due to the temperature quench. Thus, adapting Ref.~\cite{2016_Huang}, the BW model obeys the Relaxation-Plateau-Relaxation-Finite Time Scaling-Adiabatic scenario, which we expect, in the infinite size and $\tau_Q \rightarrow \infty$ limit to translate to Relaxation-Adiabatic-Impulse-Adiabatic.
\end{itemize}

However, one question still remains: \textit{ why} does a plateau form for Fig.~\ref{abheat}a when the quench is abrupt enough in the first place? This phenomenon could be explained on the basis of the following observation:
Assume for a moment that a longitudinal field is added to the BW model, Eq.~\ref{BWmodel} \cite{2005_Martinos,2011_Velonakis,2013_Velonakis}. The phase diagram of the BW model in a longitudinal field is known to possess a frustrated phase, similar to the one obtained for the triangular plaquette model (TPM) in a field \cite{2024_Sfairopoulos} and system sizes multiples of three (note that the difference of the two lies to the order of the phase transition, which for the thermal phase diagram of the BW model in a field is continuous, while for the \textit{quantum} phase diagram of the TPM first-order). For both cases, a first-order phase transition is encountered as the longitudinal field is tuned from negative to positive longitudinal fields for $T < T_c$. For our case for the BW model, we can thus infer that the BW model (without a field and the temperature lower than the critical point) lives on the coexistence of the frustrated and a trivial phase. As a result, the frustration of the negative longitudinal field phase still impacts the BW model and, we argue, that it is the reason for the metastability that we observe.

\section{Summary and discussion}{\label{conclusions}}
\label{summary}
In summary, this paper has investigated the dynamical properties of the Baxter-Wu model under linear driving, encompassing both linear cooling and heating processes. In both scenarios, the EDD in the critical region matches the predictions of the KZ mechanism. After exiting the impulse regime during linear cooling, the system undergoes a decay process with a power-law form with exponent significantly different from the exponent observed in the coarsening of the instantaneous quench. After exiting the impulse regime during the linear heating, the system does not directly enter the adiabatic regime, but first passes through a short crossover region characterized by the exponential decay of the EDD. We have also studied the dynamical process of heating from an ordered initial state that is not the ground state, and we observed a dynamical behavior that markedly deviates from the predictions of the KZ mechanism. 
We argued that this discrepancy comes from the coexistence of the frustrated phase and the trivial phase from the equilibrium phase diagram of the BW model.

Our work indicates that the dynamical behavior after exiting the impulse regime, regardless of whether it involves cooling or heating, merits further in-depth exploration. In particular, the scaling behavior of the EDD at the endpoint of the cooling is highly worthy of further investigation in other models with similar behavior, classical and quantum. The anomalous heating scenario we investigated actually reveals that, when the system is heated from a specific nonequilibrium state and crosses the critical point, its dynamical behavior may not necessarily follow from the predictions of the KZ mechanism. Recent research focusing on the Ising model has also shown that the choice of initial state affects the KZ mechanism and the coarsening dynamics \cite{Ising3}. This question also deserves extensive research in other models, especially those where the free energy exhibits complex local minima, such as frustrated spin models and spin glasses \cite{2017_Xu,2018_Xu,2022_Reichhardt}.

\section*{Acknowledgments}

We thank Jacek Dziarmaga and Shuai Yin for helpful discussions and communications.
C. D. is supported by the National Science Foundation of China (NSFC) under Grant Number 11975024;
WZ acknowledges Shanxi Province Science Foundation under Grant Number 202303021221029. 
KS acknowledges financial support from EPSRC Grant No.\ EP/V031201/1.

\bibliography{references}
\bibliographystyle{apsrev4-2}

\end{document}